# BioStatFlow, a statistical analysis workflow for "omics" data

Daniel Jacob[1,2], Catherine Deborde[1,2], Annick Moing[1,2]

[1] INRAE, Univ. Bordeaux, UMR1332 Biologie du Fruit et Pathologie, Centre INRAE de Nouvelle Aquitaine - Bordeaux, 71 av Edouard Bourlaux, F-33140 Villenave d'Ornon, France

[2] PMB-Metabolome, INRAE, 2018, Bordeaux Metabolome Facility (doi.org/10.15454/1.5572412770331912E12), MetaboHUB, PHENOME, IBVM, Centre INRAE de Nouvelle Aquitaine - Bordeaux, 71 av Edouard Bourlaux, F-33140 Villenave d'Ornon, France

Corresponding author: Daniel Jacob, daniel.jacob@ inrae.fr

# Keywords

## Abstract

BioStatFlow is a free web application, useful to facilitate the performance of statistical analyses of "omics", including metabolomics, data using R packages. It is a fast and easy on-line tool for biologists who are not experts in univariate and multivariate statistics, do not have time to learn R language, and only have basic notions in biostatistics. It guides the biologist through the different steps of a statistical workflow, from data normalization and imputation of missing data to univariate and multivariate analyses. It also includes tools to reconstruct and visualize networks based on correlations. All outputs are easily saved in a session or downloaded. New analytical modules can be easily included upon request.

BioStatFlow is available online: http://biostatflow.org

## Method summary

Here we present a free web application for routine univariate or multivariate statistical analyses of omics data for students or confirmed biologists who are not experts in statistics and do not have time to learn R language.

Statistical tools are needed in order to generate information from omics experiments. A range of applications to be run locally exists from vendors, and several initiatives have developed open and free web applications such as MultiExperiment Viewer (1). Typically, an experiment is organized to ensure that the right type of data will be produced to answer questions of interest as efficiently as possible. These specific questions must be clearly identified before carrying out this experiment. In particular, this implies defining the experimental factors. Such experimental factors are controlled independent variables the levels of which are set by the experimenter such as treatment (e.g. control vs. stress), genotype (*e.g.* wild-type vs. mutant), or the course of time (*e.g.* developmental stage).

Omics experiments yield amounts of data too large to be interpretable by a human eye. A combination of multivariate and univariate data analyses (2, 3) are therefore essential to extract and visualize the information of interest. When biologists gain basic knowledge about the statistics employed, they can contribute to evaluate their experimental design and results by themselves, and/or in interaction with statistics experts. However, there is still a lack of useful, fast and easy statistical tools on-line for who is not an expert. BioStatFlow, based on embedded R scripts, and guiding the student or confirmed biologist along the different steps of a statistical workflow, has been developed to meet this need. It can also be used as an easy tool for teaching or training basic knowledge about biostatistics.

R is a very powerful language, largely used for statistical computing (4). It runs on all important platforms and provides a range of useful specialized modules and utilities that are shared to the community (https://cran.r-project.org). However a basic knowledge of the R language and the different application conditions and steps of a statistical workflow are needed. Therefore, the development of web applications embedding R scripts for biologists remains useful as it contributes to spare time for training and allow performing a range of complementary analyses in a very short time. The biologist can easily perform a first set of basic statistical analyses and after this interact with a statistics expert.

BioStatFlow has been designed to execute statistical analyses sequentially, i.e. a linear chain of statistical processing, so-called "workflow". The default workflow is based on a set of use-cases mainly about metabolomics (2, 5). It gathers a set of univariate and multivariate methods (4, 6), to provide a complete view of data. Indeed, as recently discussed in (3), multivariate methods make use of covariances or correlations which reflect the extent of the relationships among the variables, whereas univariate methods focus solely on the mean and the variance of a single variable.

BioStatFlow also helps disseminate the results of statistical analyzes by saving them in a persistent session so that they can be fully restored. One can thus provide the session identifier when publishing results, by communicating the URL based on the template "biostatflow.org/view/<sessionID>". Example of session: http://biostatflow.org/view/S35065

The different steps of BioStatFlow follow a typical workflow (Fig. 1). After uploading the data file (csv format), a set of analyses is first proposed as a static sequence in order to normalize the dataset. At this stage, users have to follow the sequence order. When the levels of several analytical variables (features) could not be determined for all samples or that different experiments need to be compared, missing value estimation and data scaling are helpful pre-processing steps. Then, users can choose any additional method depending on the dataset and the corresponding experimental design (i.e. factors), in order to *i*) visualize the whole data, *ii*) reveal biomarkers, *iii*) analyze interactions between factors, *iv*) discriminate groups, and so on. Initial data filtering using individual names or factor levels can be used to rapidly perform analyses on a subset of the uploaded data matrix.

The input to each step takes the output of previous step. If a statistical treatment generates a data table (matrix) as an output, it will be used as input for the next step. Otherwise, if the treatment only generates results (texts and images) but does not change the input array, this latter will be directly taken as output. Each statistical treatment step has been written as an R script (most common) or as a PERL script, embedding binary tools. Examples of univariate and multivariate output results are shown Figure 2. All these outputs can be easily downloaded from a zip file or saved using a created session.

Feedback from users can be sent to the developer in order to improve the user-friendliness of the application or complement the online tutorial. Additional statistical methods and the corresponding recent and validated R packages can be easily included in the analytical workflow, and even new workflows could be created, when they are of interest for the mining of omics data including metabolomics data.

# Author contributions

DJ developed and implemented the application and wrote the manuscript. CD and AM tested the application and contributed to manuscript writing.

# Acknowledgements

We thank MetaboHUB (ANR-11-INBS-0010 project)

# Competing interests

The authors declare no competing interests.

# Figure legends

**Figure 1.** The default workflow implemented in BioStatFlow, from data normalization to data univariate or multivariate analysis. ANOVA, variance analysis; ICA, independent component analysis; PCA, principal component analysis; PLS-DA, partial least square discriminant analysis, HCA, hierarchical clustering analysis.

**Figure 2.** Examples of univariate and multivariate output results produced by BioStatFlow, on a proton NMR metabolomics profiling dataset of tomato (7).

Fig. 1

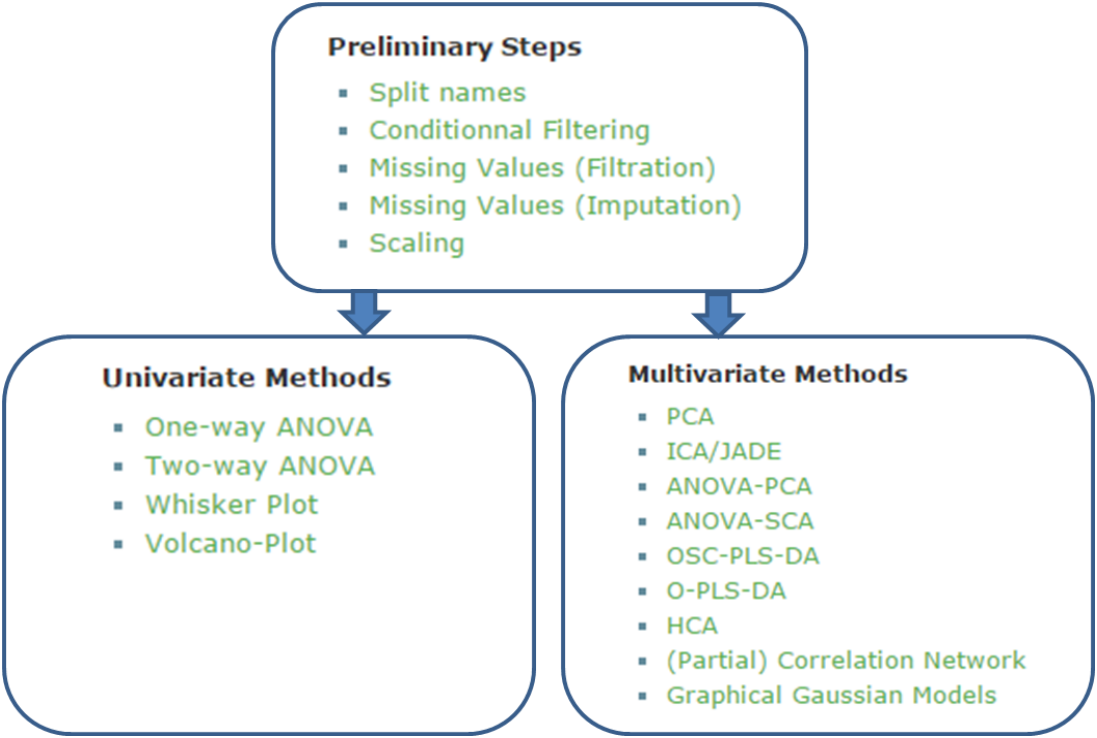

Fig. 2

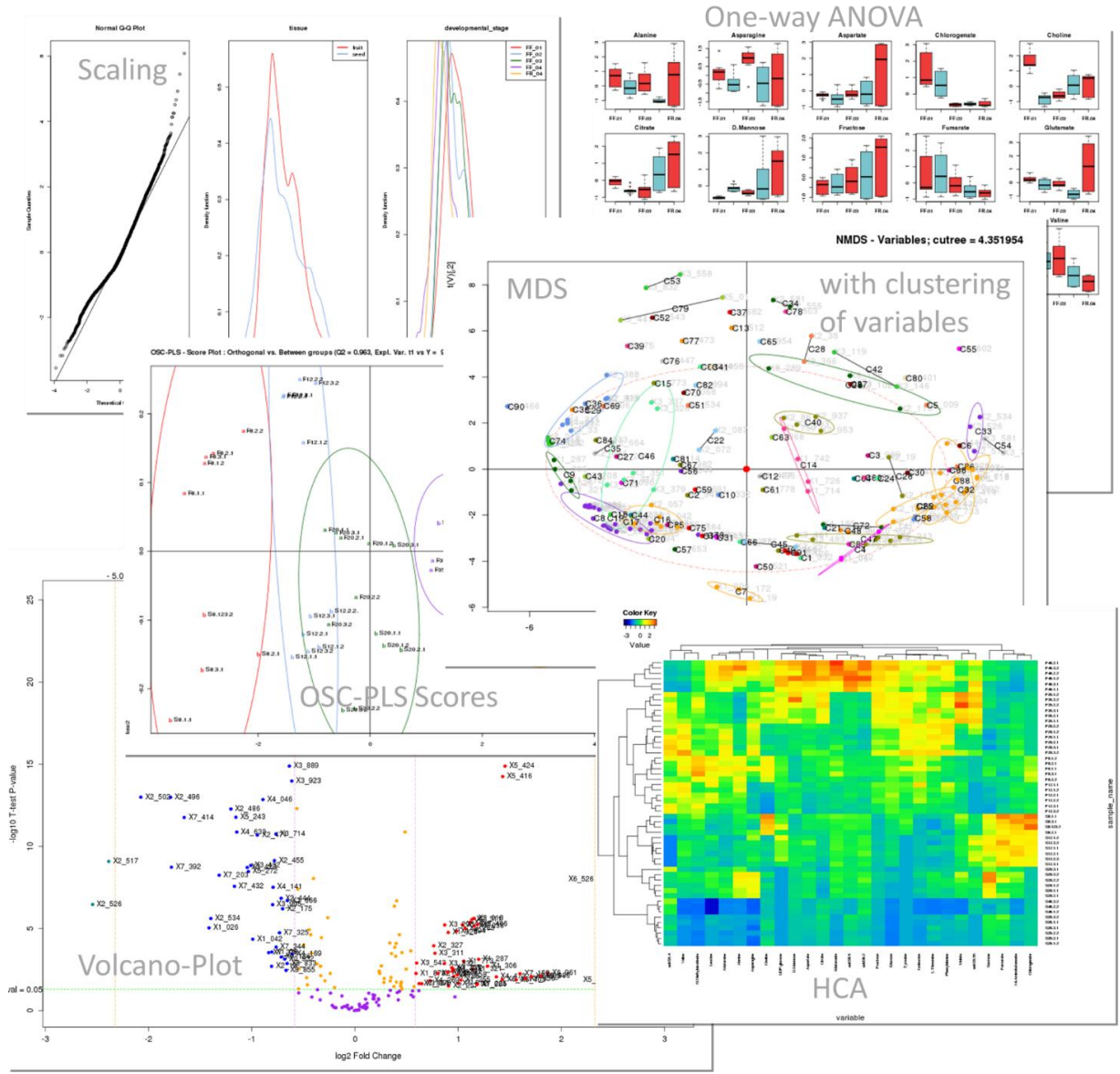